\documentclass[sn-mathphys]{sn-jnl}

\usepackage{geometry}
\usepackage{bm}
\usepackage{amsmath,amssymb,amsfonts}
\usepackage{graphicx}
\usepackage{textcomp}
\usepackage{subcaption}
\usepackage{comment}
\usepackage{blkarray}
\PassOptionsToPackage{titletoc}{appendix}

\theoremstyle{thmstyleone}%
\theoremstyle{thmstyletwo}%

\theoremstyle{thmstylethree}%

\begin{document}
\newgeometry{top=-1cm, left=1.5cm, right=1.5cm}

\font\titlefont=cmr12 at 16pt
\font\authorfont=cmr12 at 10pt
\font\affilfont=cmr12 at 10pt

\title[Financial Portfolio Optimization: a novel QUBO Formulation]{\titlefont Diversifying Investments and Maximizing Sharpe Ratio: a novel QUBO formulation}



\author[1,3]{\authorfont \fnm{Mirko} \sur{Mattesi}}
\author[1]{\authorfont \fnm{Luca} \sur{Asproni*}}
\author[2]{\authorfont \fnm{Christian} \sur{Mattia}}
\author[1]{\authorfont \fnm{Simone} \sur{Tufano}}
\author[2]{\\\authorfont \fnm{Giacomo} \sur{Ranieri}}
\author[1]{\authorfont \fnm{Davide} \sur{Caputo}}
\author[2]{\authorfont \fnm{Davide} \sur{Corbelletto} }

\affil[1]{ \affilfont \orgname{Data Reply S.r.l.}, \orgaddress{\street{corso Francia, 110}, \city{Torino}, \postcode{10143}, \country{Italy}}}

\affil[2]{ \affilfont \orgname{Intesa Sanpaolo S.p.A.}, \orgaddress{\street{piazza San Carlo, 156}, \city{Torino}, \postcode{10121}, \country{Italy}}}

\affil[3]{ \affilfont \orgname{Department of Mathematical Sciences, Politecnico di Torino}, \orgaddress{\street{Corso Duca degli Abruzzi, 24}, \city{Torino}, \postcode{10129}, \country{Italy}}}

\abstract{}
The Portfolio Optimization task has long been studied in the Financial Services literature as a procedure to identify the basket of assets that satisfy desired conditions on the expected return and the associated risk. A well-known approach to tackle this task is the maximization of the Sharpe Ratio, achievable with a problem reformulation as Quadratic Programming. While the sole Sharpe Ratio could be efficiently optimized via classical solvers, in business scenarios it is common that multiple additional needs arise, which have to be integrated in the optimization model as either new constraints or objective function terms. Then, in general, the problem may become non-convex and hence could potentially be not efficiently solvable via classical techniques anymore.\\
One example of such additional objective function term consists of maximizing a diversification measure penalizing portfolios holding significant portions of investments on assets belonging to the same sector, while favouring solutions that diversify over multiple sectors. The problem of optimizing both the Sharpe Ratio and a diversification term can be mapped to a QUBO and be solved via quantum annealing devices or Hybrid Computing approaches, which are expected to find high quality solutions. 
We propose a new QUBO formulation for the task described and provide the mathematical details and required assumptions, showing the ease of modeling the optimization as QUBO against the effort that would be required by classical strategies. We derive results via the available QUBO solvers, as well as discussing the behaviour of Hybrid approaches to tackle large scale problems in the near term. We finally elaborate on the results showing the trade-off between the observed values of the portfolio's Sharpe Ratio and diversification, as a natural consequence of solving a multi-objective optimization problem.
\footnote[0]{*Corresponding author. E-mail: l.asproni@reply.it}




\maketitle
\restoregeometry

\section{Introduction}\label{introduction}

Portfolio optimization plays a pivotal role in the financial industry. 
Banks, insurance companies, and hedge funds exploit the theory of modern portfolio formulated by Harry Markowitz, which earned him the Nobel Prize \cite{Markowitz1952}. 
In Portfolio Optimization, we are given a set of assets to choose from. Although some portfolios are handcrafted uniquely according to the experience of financial advisors, a software market has emerged and continues to grow in tandem with technological advancements. By considering the evolving behavior of these assets' values, we can estimate the expected return (financial gain) and its volatility (risk). We then utilize this information to construct an optimization problem that is built on the maximization of returns and minimization of the risk. 
Many of such objective functions have been engineered, all aiming to construct the best portfolio, \textit{Sharpe Ratio} has emerged as one of the main indicators of portfolio quality \cite{sharpe1998sharpe}. The Sharpe Ratio is a measure of an investment's performance, calculated as the ratio between the portfolio's expected return and its risk, computed as standard deviation. In its simplest form, the resulting
optimization problem is convex, thus efficient to solve on a classical device.
 
Portfolio optimization has been approached by different means, including linear programming \cite{mansini2014twenty}, quadratic programming \cite{best2000quadratic}, semidefinite programming \cite{fonseca2011semidefinite}, meta-heuristics \cite{erwin2023meta}, deep learning \cite{heaton2017deep}, and reinforcement learning \cite{almahdi2017adaptive}.

In business scenarios, the Sharpe Ratio is often considered a fundamental, yet not fully exhaustive, measure to evaluate financial portfolios. Other characteristics are taken into account, based on the needs that the portfolio is required to meet. One such example is the degree of diversification: investments spread across multiple sectors and industries are preferred for risk-averse profiles, which can often be the case of large financial institutions. Modeling additional business needs into an optimization problem may lead to non-trivial tasks, rising the demand for efficient strategies able to find high-quality solutions.
These needs lead to different objective functions,
which are based on the Sharpe Ratio optimization but make the resulting problem not necessarily convex anymore. A common business directive is the \emph{diversification} designed as the preference towards
solutions that allocate budget on as weakly correlated (or anticorrelated) families of assets as possible. Stocks of different companies from the same industry, for example real estate, may be somewhat linked and the resulting portfolio not appealing to an investor.


Recently, the improvement in computational capabilities promised by quantum computing has generated large interest in the financial sector \cite{Herman2023}. One of the first applications of quantum annealers has been proposed by authors in \cite{venturelli}, who built a Quadratic Unconstrained Binary Optimization
(QUBO, \cite{Asproni2020, incudini2022computing}) formulation for Portfolio Optimization.
This formulation is tailored to run with the limited capabilities of the quantum hardware
available at the time of writing. This has resulted in modeling Sharpe maximization as a discrete problem, where each asset is either selected or not, and so, breaking the convexity of continuous model formulation, it makes the problem nonconvex. 

However, with the hardware constantly improving, we are interested in developing methods for a wider scope \cite{king2023quantum}.

In this work, we propose a novel QUBO formulation that preserves the Sharpe Ratio and adds a diversification term to the optimization problem. Firstly, we modify the approach in \cite{venturelli} to go beyond the all-or-nothing selection approach of each asset, allowing the selection of a linear
combination of investments on all assets with arbitrarily large precision: the more precise the linear combination becomes, the more binary variables are needed and the formulation itself becomes more convex, up to the limit case of infinite precision which is
convex. The possible convexity of such a formulation, which at this point can be solved efficiently by classical means, do not undermine the importance of the work because the formulation can serves as a baseline to add personalized, business-dependent constraint.
Secondly, we propose the diversification as one of the possible terms that might be taken into consideration. The presence of a diversification
objective function item which has the aim to penalize,
but potentially not completely rule out, the investments on assets belonging to the same market sector, guaranteeing a portfolio diversified over multiple sectors up to an arbitrary degree. The assets were then grouped according to the information reported by the provider into the following 12 sectors: Basic Materials, Communication Services, Consumer Cyclical, Consumer Defensive, Energy, Financial Services, Healthcare, Industrials, Real Estate, Technologies, Utilities. This solution maximizes the expected return and, at
the same time, minimizes the risk of loss given by potential market crashes in individual sectors. We show how to integrate such a term on the baseline QUBO formulation, proving that the diversification maximization leads to a nonconvex problem.



We test our approaches on the classical \textit{qbsolv} algorithm and on the \textit{D-Wave Leap hybrid} classical-quantum solver. In particular, the choice of such solvers allows us to abstract from all the hardware detail that we need to consider when executing our instances on quantum annealers. The implication of porting our approach on noisy, small-scale quantum annealers is discussed. Alternative approaches in solving the Portfolio Optimization on quantum computers, although not exhaustive for the task that we face, are shown in \cite{PhysRevApplied.15.014012, Black7, kerenidis2019quantum, rebentrost2018quantum}.

The paper is structured as follows. Section \ref{sec:preliminaries} is devoted to reviewing the background on the topic. Section \ref{sec:formulation} describes the fine-grained QUBO formulations for Sharpe maximization. Section \ref{sec:diversif} shows the chosen modeling strategy for the diversification term. In Section \ref{sec:experiments} we reports our experiments with various solvers and we discuss the results. Finally, Section \ref{sec:conclusions} concludes our work and outlines future directions.

\section{Preliminaries}\label{sec:preliminaries}

In this section, we briefly introduce the necessary background to render our work self-contained. For further details, one can refer to \cite{glover2018tutorial} for the definition of effective QUBO formulations and to \cite{rajak2023quantum} for quantum annealing.

\subsection{Notation}
The sets of real, complex, and Boolean values are denoted with $\mathbb{R}, \mathbb{C}, \mathbb{B} = \{0, 1\}$, respectively. Scalar variables and constants are denoted with lowercase alphabetic characters and lowercase Greek characters respectively, e.g. $x$ and $\lambda$. Vectorial values are denoted with lowercase bold characters, e.g. $\bm{x}$, and are intended to be column vectors. Matrix and operators are denoted with uppercase alphabetic characters, e.g. $Q$. The notation $\sigma_x, \sigma_y, \sigma_z$ denotes the Pauli matrices, with the optional apex indicating the qubits on which the operator acts. 

\subsection{Quadratic Unconstrained Binary Optimization}
The Quadratic Unconstrained Binary Optimization (QUBO) is an NP-hard combinatorial optimization problem defined as follows: given a set of $n$ Boolean variables $\bm{x} \in \mathbb{B}^n$ and a $n\times n$ matrix $Q$ - typically in upper-triangular form, of real values, the problem consists of finding the value
\begin{equation}
    \bar{\bm{x}} = \arg\min_{\bm{x}} \bm{x}^\top Q \bm{x},
\end{equation}
or equivalently
\begin{equation}\label{eq:qubo}
    \bar{\bm{x}} = \arg\min_{\bm{x}} \sum_{i=0}^{n-1} \sum_{j=i}^{n-1} x_i Q_{ij} x_j.
\end{equation}

Despite the simplicity of this approach, its NP-hardness guarantees the possibility of expressing a vast class of problems. It is notable that we can express problems over integers and rational variables using different encodings. For example, a positive integer $y$ can be defined over an $m$-bit variable $\bm{b}$ through the binary encoding $y \equiv \sum_{i=0}^{m-1} 2^i b_i$ or through the unary encoding $y \equiv \sum_{i=0}^{m-1} b_i$. The use of different encodings will lead to diverse QUBO formulation and their efficacy has to be assessed on a per-case basis. The same reasoning applies to differently structured data, e.g. fixed point rational variables. 

Constraints can be expressed implicitly representing the QUBO in  \ref{eq:qubo} as a linear combination of the main objective function and other terms, each corresponding to one constraint. In \cite{dattani2019quadratization} it is shown how to design different kinds of constraint terms. 


\subsection{Quantum Annealing}
Quantum annealing is a heuristic optimization procedure proposed in \cite{ray1989sherrington}. It is carried on by initializing a quantum mechanical system in a superposition of all candidate solutions and evolves according to the Schr\"odinger equation under a time-varying hamiltonian:
\begin{equation}
H(t) = \Gamma(t) H_0 + \mathcal{J}(t) H_f
\end{equation}
Here, $t \in \mathbb{R}_{\geq 0}$ represents the time, $H_0 = -\sum_i \sigma_x^{(i)}$ is the hamiltonian of the system at initialization, and $H_f$ corresponds to the problem formulation to be minimized. According to the adiabatic theorem \cite{farhi2001quantum}, if the system evolves slowly enough, it remains in its ground state throughout the entire evolution. The evolution is controlled by the given schedule, specified by $\Gamma$ and $\mathcal{J}$. The hamiltonian $H_f$ takes the form:
\begin{equation}
H_f = \sum_{i=0}^{n-1} h_i \sigma_z^{(i)} + \sum_{i=0}^{n-1} \sum_{j=i+1}^{n-1} J_{ij} \sigma_z^{(i)} \sigma_z^{(j)}
\end{equation}

where $h_i$ is called \textit{bias} and represents the strength that leads the corresponding $S_i$ variable to take either value $-1$ or $1$,  $J_{ij}$ is called \textit{coupler} and encodes the relationship between pairs of variables $(S_i, S_j)$ and both are real-valued parameters. The measurement results in a vector $\bm{S}$ of spin variables $S_i \in \{{\pm 1}\}$. It is worth noting that the QUBO formulation can be naturally expressed as an Ising hamiltonian, where a change of variables $S_i \leftrightarrow 2 x_i - 1$ establishes the equivalence.

Recently, several companies have developed large-scale quantum annealers, which are specialized Quantum Computers designed for solving Ising hamiltonians. Examples of such systems include D-Wave \cite{king2023quantum} and Qilimangiaro \cite{canivell2021startup}. The availability of quantum annealers is crucial to address the class of problems that can be solved through a QUBO model, aiming to achieve an advantage over other techniques in terms of both solution quality and computational speed.

\subsection{Combinatorial Optimization Techniques on Quantum Computers}
The development of combinatorial optimization algorithms for Quantum Computers is of paramount interest. Most combinatorial optimization problems are NP-hard, which informally means they are at least as hard to solve as the most challenging NP problem. It is widely believed, based on reasonable computational complexity assumptions \cite{aaronson2005guest}, that neither classical nor Quantum Computers can efficiently solve NP-hard optimization problems. However, significant theoretical speedup has been proven, and various techniques, including quantum annealing, have shown promise in providing improved solutions for certain classes of problems. 

Regarding quantum annealing, the adiabatic theorem does not guarantee efficient convergence to the global optimum, as the time required to evolve the system may be exponential in the size of the problem instances. In comparison to the simulated annealing algorithm, authors in \cite{farhi2002quantum} have proven that quantum annealing can leverage quantum tunneling effects to escape local minima, where simulated annealing would require exponential time to escape. On the other hand, \cite{van2001powerful} presents arguments in favor of simulated annealing. More recently, \cite{liu2023quantum} have identified specific characteristics that can give quantum annealing an advantage over classical techniques, such as landscapes with many local minima separated by high but thin barriers. 

Different possibilities are suggested by quantum algorithms for semidefinite programming \cite{brandao2022faster}, which can offer significant speedup over the current classical solution but require fault-tolerant quantum hardware. Authors in \cite{augustino2023solving} have shown how to formulate semidefinite relaxations of QUBO problems.

\subsection{Sharpe Ratio Maximization in Portfolio Optimization}
Portfolio Optimization is a family of combinatorial optimization problems where the objective is to find an optimal allocation of weights for different assets. The goal is to simultaneously maximize the expected return or profit from the portfolio and minimize the associated risk. A potential approach for finding a suitable portfolio is the maximization of the Sharpe Ratio. For an in-depth treatment of Portfolio Optimization techniques one can refer to \cite{cornuejols_2006}.

Formally, given a set of $n$ assets, let $\bm{\mu} = (\mu_1, ..., \mu_n)$ be the vector of expected returns of such assets and $\bm{w} = (w_1, ..., w_n)$ be a weight vector such that $\sum_{i=1}^n w_i = 1$. The total expected return of the portfolio is calculated as the weighted sum of the expected returns of each asset, i.e. $\bm{w}^T \bm{\mu}.$

The \textit{risk} or volatility of the portfolio is quantified by the standard deviation, denoted as $\sigma$, which is the square root of the portfolio variance. The portfolio variance is computed as the quadratic form of the weight vector and the covariance matrix $\Sigma$, i.e. $\bm{w}^\top \Sigma \bm{w}$.

The Sharpe Ratio, denoted by $S$, which does not consider a risk-free rate, is defined as ratio between the expected return of a portfolio and the square root of its variance. Then, the problem that we solve, which we will denote as \textit{Max-Sharpe} problem from now on, is formulated as
\begin{align}\label{maxsharpe}
    \max & \frac{\bm{w}^T\bm{\mu}}{\sqrt{\bm{w}^T\Sigma \bm{w}}}\\ 
    \text{s.t. } & \sum_{i=1}^n w_i = 1\\
    & w_i \geq 0 \quad \forall i=1,\cdots,n
\end{align}

The presence of additional constraint, such as regularization or sparsity
conditions, may result in a problem more challenging to solve.

\subsection{Diversification of investments over multiple sectors}\label{subsec_diversif}

While we model the Sharpe Ratio maximization as the first building block of our task, in some business scenarios multiple additional needs arise, which can be encoded as constraints in the optimization problem or as additional terms in the objective function. The presence of these terms could make the problem no longer convex and therefore not efficiently solvable via classical techniques.

The need for diversification over multiple sectors, when modeled as objective function item, leads to a nonlinear and, in general, nonconvex optimization problem. Let's assume that we have a matrix $A$ with 5 assets $(x_1, ..., x_5)$ that belong to 3 sectors $(s_1, s_2, s_3)$ that are defined in the following way: 
\[ A =
\begin{blockarray}{cccccc}
 & x_1 & x_2 & x_3 & x_4 & x_5 \\
\begin{block}{c(ccccc)}
  x_1 & -0.5 & 1 & 0 & 0 & 0  \\
  x_2 & 1 & -0.5 & 0 & 0 & 0  \\
  x_3 & 0 & 0 & -0.5 & 1 & 0  \\
  x_4 & 0 & 0 & 1 & -0.5 & 0  \\
  x_5 & 0 & 0 & 0 & 0 & -0.5  \\
\end{block}
\end{blockarray}
 \]
As can be seen from $A$, $x_1$ and $x_2$ belong to $s_1$, whereas $x_3$ and $x_4$ belong to $s_2$, and $x_5$ is the only asset that belongs to $s_3$.
We formulate the term as a penalization of investments on assets belonging to the same sector, while incentivizing the allocation of capital on the individual assets

\begin{align}\label{eq:diversif}
    \min \quad \bm{w}^T \bm{f} + \bm{w}^TD\bm{w}
\end{align}

where $f_i < 0$ $\forall i=1, \cdots, n$ is the vector of components that drive the solution to invest a positive quantity, $D\in \mathbb{B}^{n \times n}$ is the matrix incorporating the information on the assets sectors and defined as follows

\begin{align}\label{diversifmatrix}
    D_{ij} = \begin{cases}
    1 & \text{asset i and j belong to the same sector}\\
    0 & \text{otherwise}
    \end{cases},
\end{align}

and $w_i$ $\forall i=1, \cdots, n$ is the investment on asset $i$.

The rationale behind this formulation lies in the preference of having a term that penalizes investments on assets belonging to the same sector and, at the same time, by favoring non null allocation on the assets, drives the optimization towards a solution that distributes investments as equally as possible over all sectors.

\section{QUBO Formulation for Sharpe Ratio Optimization}\label{sec:formulation}

In \cite{venturelli}, the authors proposed the standard de facto QUBO formulation for Sharpe Ratio maximization. Given a portfolio of $n$ assets, the QUBO is defined as follows:
\begin{equation}\label{eq:boolean_formulation}
\overline{\bm{q}} = \arg\min_{\bm{q}} - \sum_{i=1}^n a_i q_i + \sum_{i=1}^{n} \sum_{j=i+1}^n b_{ij} q_i q_j,
\end{equation}
where $q_i$ is a binary variable representing whether the $i$-th asset is selected ($q_i = 1$) or not ($q_i = 0$), $a_i$ represents the expected risk-adjusted return ($a_i = \mu_i / \sigma_i$), and $b_{ij}$ represents the diversification penalties and rewards, corresponding to the correlation between assets $i$ and $j$. 

Optionally, for $\lambda \in \mathbb{R}_{\ge 0}$, the formulation with an additional term formulated as
$$\lambda \left(M - \sum_{i=1}^n q_i\right)^2$$
would reward solutions having $M$ selected assets.

Furthermore, the authors denote the need of grouping the values of $a_i$ into buckets of 11 evenly spread ranges, and $b_{ij}$ into buckets of non-evenly spread ranges. Such a step likely leads to a non positive definite correlation matrix, making the problem more complex for classical solvers. 


\subsection{Sharpe Ratio Proxy formulation}

One important aspect to note in this approach is the coarse-grained selection, where each asset is either selected or not.

In order to solve a case with a wider scope where we are interested in finding optimal quantity of investments over the assets, we propose the introduction of fractional weights. To do so, we substitute each $1$-bit weight $q_i$ with a $p$-bit vector $\bm{x}_i$, encoded according to the formula:
\begin{equation}
\sum_{\ell=0}^{p-1} \frac{2^\ell}{500} x_{i\ell}.
\end{equation}
This encoding, with a chosen precision of $p=9$, allows for discretizing the range $[0,1]$ with a granularity of $0.002$. This means that the minimum investment in a single asset is $0.2\%$. We will denote the discretization constant as $d_\ell$, which, in this example, has been set to $2^\ell/500$.

The overall formulation is defined on the binary matrix of variables $\bm{x} \in \mathbb{B}^{n \times p}$, with a suggested value of $p=9$, in the form
\begin{equation}\label{eq:fractional_formulation}
Q(\bm{x}) = \lambda_0 H_0(\bm{x}) + \lambda_1 H_1(\bm{x}),
\end{equation}
where
\begin{equation}\label{sharpe_proxy_obj}\begin{split}\begin{aligned}
H_0(\bm{x}) =
\sum_{i=1}^n a_i \left(\sum_{\ell=0}^{p-1} d_\ell x_{i\ell}\right) + \\
\sum_{i=1}^n\sum_{j=i+1}^n b_{ij} \left(\sum_{\ell=0}^{p-1} d_\ell x_{i\ell}\right) \left(\sum_{\ell=0}^{p-1} d_\ell x_{j\ell}\right)
\end{aligned}\end{split}\end{equation}

is the main objective function to minimize, and
\begin{equation}\label{sharpe_proxy_constr}
H_1(\bm{x}) = \left(\sum_{i=1}^n \sum_{\ell=0}^p d_\ell x_{i\ell} - 1\right)^2
\end{equation}
is the reward term for solutions that satisfy the constraint $\sum_i w_i = 1$. Note that our formulation only has a linear overhead in the number of variables required.

The constants $\lambda_0$ and $\lambda_1$ are hyperparameters that can be chosen through educated guesses, grid-search, or black-box optimization methods such as Bayesian optimization \cite{snoek2012practical}. To ensure the satisfaction of the constraint, we should impose $\lambda_1 \gg \lambda_0$, which is convenient for classical devices with little numerical error on the encoding, but less convenient on a quantum annealer.

However, it is important to note that optimizing these formulations does not directly equate to maximizing the Sharpe Ratio as it is mathematically defined in \ref{maxsharpe}. It is desirable to construct an objective function that accurately reflects the definition of the Sharpe Ratio.
Namely, defining for simplicity the variables $z_m = x_{ik}$ $\forall m=1,\cdots,p*n$ and knowing that $p=9$:
$$
\sum_{i=1}^{9n} \sum_{j=i+1}^{9n} Q_{ij} z_i z_j \neq \frac{\bm{\mu}^T\bm{w}}{\sqrt{\bm{w}^T\Sigma \bm{w}}}
$$

This is due to the lack of information related to the Sharpe Ratio of the overall portfolio. Instead, only individual contributes of each asset's Sharpe Ratio and a correlation factor between assets are taken into account.
Therefore, we propose a novel formulation that aims to maintain faithfulness to the original definition of the Sharpe Ratio.

\subsection{Proposed Sharpe Ratio Formulation}\label{proposed_form}

Maximizing the Sharpe Ratio as defined in \ref{maxsharpe} is a nonlinear optimization problem. Authors in \cite{cornuejols_2006} provide a quadratic reformulation of the problem by introducing a change of variables, namely:

\begin{align}\label{reformulated}
    \min & \quad \bm{y}^T\Sigma \bm{y}\\ 
    \text{s.t. } & \left(\bm{\mu} - r_f \bm{e} \right)^T \bm{y} = 1\\
    & (\bm{y}, k)\in \mathcal{Y}
\end{align}

where \begin{equation}
    \mathcal{Y} := \left\{(\bm{y}, k) \mid \bm{y} \in \mathbb{R}^n, k\in \mathbb{R}_{> 0}, \bm{w}=\frac{\bm{y}}{k}\in \mathcal{W}\right\} \cup (0,0),
\end{equation}
$r_f$ is the risk-free rate, $\bm{e}$ is the vector of ones, $\mathcal{W}$ is the set of feasible portfolios such that $\bm{e}^T\bm{w} = 1 \quad \forall \bm{w}\in\mathcal{W}$ and under the assumption that $\exists \hat{\bm{w}} \in \mathcal{W}$ $ \vert$ $ \bm{\mu}^T\hat{\bm{w}} > r_f$. This way, given the optimal solution $(\overline{\bm{y}}, \overline{k})$, it is possible to retrieve the optimal portfolio allocation as $\overline{\bm{w}} = \frac{\overline{\bm{y}}}{\overline{k}}$.

In order to build our QUBO formulation for the problem above, we consider $r_f = 0$, we identify the set of feasible portfolios $\mathcal{W} = \{\bm{w} \in \mathbb{R}^n \vert \bm{e}^T\bm{w}=1, w_i\geq0 \quad \forall i=1, \cdots, n\}$, meaning that we do not introduce additional linear constraints and finally we assume $\mu_i > 0 \quad \forall i = 1\cdots, n$.

We rely on the latter in order to define the discretization of the new variables. This assumption, along with the constraint $\bm{\mu}^T \bm{y} = 1$, allows to find an upper bound for the $y$ variables equal to $\frac{1}{\mu_{min}}$, where $\mu_{min}$ is the smallest (positive) expected return of our assets.
This consideration is backed up by the following, which holds under our assumption:
$$\bm{\mu}^T\bm{y}= 1 \implies \sum_{i=1}^N \mu_i y_i = 1 \implies y_i \leq \frac{1}{\mu_{min}} \quad \forall i=1, \cdots, n$$

The quadratic formulation of our optimization problem is as follows:

\begin{equation}\label{final_classical_problem}
\begin{aligned}
\text{min } & \bm{y}^T\Sigma \bm{y} \\
\text{s. t. } & \bm{\mu}^T\bm{y} = 1 \\
& y_i\in \left[0, \frac{1}{\mu_{min}}\right] \forall i \in 0, \cdots, n \\
\end{aligned}
\end{equation}

where $\overline{k} = \bm{e}^T\overline{\bm{y}}$ and the optimal portfolio allocation in terms of assets weights $\bm{w}$ is found as $\overline{\bm{w}} = \frac{\overline{\bm{y}}}{\overline{k}}$.

For our dataset, we find that $\mu_{min}=0.00245$, from which follows that $\frac{1}{\mu_{min}} = 408.10190$. Therefore, in our QUBO model, we discretize the quantities in the range $\left[0, 408.10190\right]$. We define the following coefficients:
\begin{equation}\label{discretiz}
c_k = 2^{k}/10 \quad \forall k=0, \cdots, p-2
\end{equation}
$$
c_{p-1} = \frac{1}{\mu_{\min}} - \sum_{k=0}^{p-2} c_k
$$
where $p=12$ allows to represent the range $\left[0, 408.10190\right]$ with a discretization step equal to $0.1$.

Finally, our proposal for a novel QUBO formulation that maximizes the Sharpe Ratio under the assumptions previously declared writes as follows:

\begin{equation}\label{eq:qubo_partial_max_sharpe}
    Q = \lambda_0 H_0 + \lambda_1 H_1
\end{equation}

where $\lambda_0$ and $\lambda_1$ are hyperparameters that must be tuned in order to find solutions that are both feasible and yield the highest value for the Sharpe Ratio,
\begin{equation}\label{proposed_obj}
H_0(\bm{x}) = \sum_{i=0}^n\sum_{j=i}^n \Sigma_{ij} \left(\sum_{k=0}^{p-1} c_k x_{ik}\right) \left(\sum_{k=0}^{p-1} c_k x_{jk}\right)    
\end{equation}
and
\begin{equation}\label{proposed_constr}
    H_1(\bm{x}) = \left(\sum_{i=0}^n \sum_{k=0}^{p-1} \mu_i c_k x_{ik} - 1\right)^2,
\end{equation}

and $\bm{x} \in \mathbb{B}^{p\cdot n}$.

\section{QUBO Formulation for Diversified Portfolio Optimization}\label{sec:diversif}

In this section, we propose the QUBO formulation for Equation \ref{eq:diversif} according to the variables defined in Equation \ref{eq:qubo_partial_max_sharpe}, we present the complete QUBO formulation for our task, comprehensive of both the Sharpe Ratio and Diversification Maximization and finally discuss the potential strategies required by classical techniques to tackle our problem, in particular the diversification term.

Building on the variable definition as in the model from Equation \ref{eq:qubo_partial_max_sharpe}, we model the QUBO term in charge of maximizing the diversification as follows:

\begin{equation}\label{eq:diversif_qubo}
    H_2(\bm{x}) = \sum_{i=0}^n \sum_{k=0}^{p-1} f_i x_{ik} + \sum_{i=0}^n \sum_{k=0}^{p-1}\sum_{j=0}^n \sum_{l=0}^{p-1} D_{ij} x_{ik} x_{jl},
\end{equation}

where $f_i$ and $D_{ij}$ are defined as in section \ref{subsec_diversif}.

Finally, the complete QUBO model is formulated as:

\begin{equation}\label{eq:qubo_final}
    Q = \lambda_0 H_0 + \lambda_1 H_1 + \lambda_2 H_2
\end{equation}

where $\lambda_2$ is an additional hyperparameter similar to $\lambda_0$ and $\lambda_1$ such that, for higher values, leads to more diversified solutions and $H_0$ and $H_1$ are as defined in Section \ref{proposed_form}.

In order to quantify the degree of diversification of different portfolios, we employ the following indicator which we call \textit{Diversification Entropy}:

\begin{equation}\label{eq:qubo_final}
    \textit{Diversification Entropy} = -\frac{\sum_{s=1}^{S}A_s \log{A_s}}{S}
\end{equation}

where $s=1, \cdots, S$ identifies the sector, $S$ is the total number of sectors and 
\begin{equation}\label{diversif_entropy}
    A_s = \sum_{\text{i s.t. asset i belongs to sector s}} w_i
\end{equation}

is the quantity of investment allocated on sector $s$. This measure takes values in $[0,1]$, $0$ corresponding to portfolios not diversified at all and $1$ indicating an equal investment over all sectors.

Classical techniques for solving a Portfolio Optimization task with a term as in Equation \ref{eq:diversif_qubo}, or even more nonconvex items modeling additional needs, may become inefficient as the problem size scales. 
Also, with a suitable change of variables, one could reformulate the objective function in order to linearize the problem and solve it through Linear Programming strategies. However, this method leads to a considerable increase in the number of variables, making it difficult for classical solvers to efficiently find satisfactory solutions for large-scale problems.
As an example, considering $N=500$ linear variables - $w_i$ as per our notation, one would need to build an additional variable for each pair of $w_i, w_j$ in the model in order to associate a penalization coefficient to assets belonging to the same sector: this leads to a total number of $124750$ additional variables that a Quadratic Programming model such as QUBO would not need. Moreover, additional constraints would be added in order to efficiently integrate the linearization variables within the problem. As the number of assets, and hence linear variables, scales, the effort to model a diversification or equivalently nonlinear term grows considerably.
With the ongoing research and development on quantum computing, which is expected to find high-quality solutions in complex scenarios, it is of paramount importance to define combinatorial optimization tasks in such a way that Quantum Computers may solve the problem.

\section{Experimental assessment}\label{sec:experiments}

In this section, we show the results of our experiments regarding two main aspects of our proposed QUBO formulation: on one hand, following the core scope of our work, we report the behaviour of the complete model as we vary the parameters influencing the Sharpe Ratio and the Diversification terms, discussing on the trade-off of these measures for the optimized portfolios; on the other hand, as an additional investigation, we evaluate the performance of our formulation for the sole Sharpe Ratio Maximization in comparison to other techniques, where the goal is to discuss the appropriateness of our model with respect to the so-called Sharpe Ratio Proxy. For the second part of our study, we consider the closeness of the Sharpe Ratio values between the portfolios derived from QUBO formulations and the one yielded by a classical optimization, which is expected to efficiently solve the problem maximizing the sole Sharpe Ratio. The benchmarking analysis is conducted on a real-world dataset of assets obtained from Yahoo! Finance. 

To solve the QUBO formulations, we employ the classical solver qbsolv and the hybrid classical-quantum solver D-Wave Leap \cite{dwavelib}. In order to obtain the solution from a classical strategy, we use \textit{PyPortfolioOpt} \cite{pypfopt}, a state-of-the-art Python3 library for Portfolio Optimization.

\begin{figure}[t!]
\captionsetup[subfigure]{justification=centering}
    \centering
      \begin{subfigure}{0.48\textwidth}
        \includegraphics[width=\textwidth]{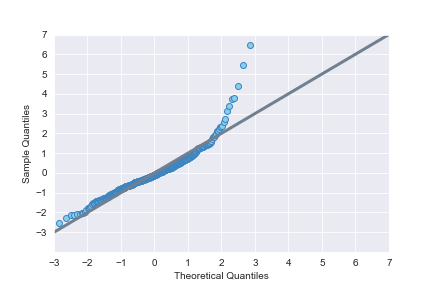}
          \caption{Simple returns}
      \end{subfigure}
      \begin{subfigure}{0.48\textwidth}
        \includegraphics[width=\textwidth]{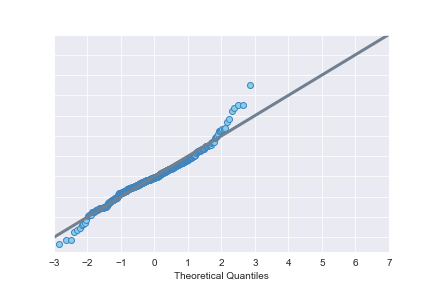}
          \caption{Log returns}
      \end{subfigure}
      \caption{Quantile-Quantile plots for simple returns and log returns}
    \label{returns_distribution}
\end{figure}

\subsection{Choice of the Solver}

The choice of solver plays a crucial role in optimization and can significantly impact the quality of the obtained solutions. To ensure a fair comparison, we utilize two different solvers: D-Wave's qbsolv and the D-Wave Leap model \textit{hybrid\_binary\_quadratic\_model\_version2}. The former is a classical technique that decomposes large QUBO matrices into smaller sub-instances, which are then solved using Tabu Search \cite{glover1993tabu}. The latter is a hybrid classical-quantum solver that combines classical and quantum optimization methods.

The use of a hybrid solver is of great importance in overcoming the current limitations of quantum hardware. Notably, the D-Wave Quantum Processing Unit (QPU) imposes restrictions on the precision at which the QUBO formulation can be encoded, requiring various adjustments and approximations at the mathematical level to compensate for these limitations. Additionally, configuring the numerous settings of a quantum annealer, such as the schedule and annealing time, is non-trivial. The hybrid solver automatically configures the optimization problem to run on classical resources and on the QPU (in our case, the Advantage QPU based on the Pegasus topology with more than 5000 qubits), without the need for manual configuration. It is important to note that the specific details of how the computing resources are utilized within the solver are not publicly disclosed.

\subsection{Dataset Specifications and Preprocessing}\label{description}

We focus on the sequence of adjusted close daily prices for the 505 assets belonging to the S\&P 500 index within the time interval of 2013 to 2020, obtained from Yahoo! Finance. To prepare the data, we follow the procedure outlined below:

\begin{enumerate}
    \item Remove null values and exclude assets whose time series contain consecutive null values, meaning null values for consecutive days.
    \item Calculate the simple returns $R_t$, defined as $R_t = \frac{{P_t - P_{t-1}}}{{P_{t-1}}}$, and the log-returns $logR_t$, defined as $logR_t = \log(1+R_t)$.
    \item Compute the sample expected returns and sample covariance matrix, annualizing the results using a frequency factor of 252, which represents the assumed number of trading days in a year.
    \item Remove assets with negative expected returns.
\end{enumerate}
The last step is not strictly necessary for a generic Portfolio Optimization strategy. However, it is an assumption for the QUBO formulation proposed in this work in order to provide the upper bound on the $\bm{y}$ variable as shown in the previous section.

After applying these steps, we are left with 460 assets in the case of simple returns and 432 assets in the case of log-returns. 
The discrepancy derives from the fact that 28 assets yield less than 1, which turns negative after logarithmic transformation.
We make the assumption that for any arbitrary asset $i$, the series $\{R_{i,t}\}_{t\geq 1}$, or equivalently the series of log-returns, follows a sequence of independent and identically distributed Gaussian random variables. This assumption allows us to use the sample mean and sample covariance matrix as estimators for the expected return and the covariance matrix of the assets, respectively. While there may be other financial time series models that could potentially be more suitable for the dataset, the exploration of such models is beyond the scope of this work. In order to have an acceptable precision we need to discretize the formulation in \ref{discretiz} using $p=12$ bits of precision for the variables.

Figure \ref{returns_distribution} shows the comparison between the distributions of simple and log returns through a Quantile-Quantile plot. Given our assumptions on the Gaussian distribution of the assets, we also perform a Shapiro test on the two distributions of returns and base our optimization on the data that shows the highest probability of following a Gaussian distribution, which leads to suggesting the use of log returns.

\subsection{Evaluation of the Solutions}
It is important to note that our main focus in this work is to provide a novel QUBO formulation to tackle a complex instance of the Portfolio Optimization problem incorporating multiple business needs. We also stress the ease of modeling quadratic and nonconvex objective function items in the QUBO model, in contrast to the effort required by classical techniques, as discussed in Section \ref{sec:diversif}. 
Our interest is then to investigate the behavior of our formulation as the impact of a diversification term on the optimized portfolio intensifies, using the Sharpe Ratio and Diversification Entropy as indicators of the solutions quality. Furthermore, we study the appropriateness of our QUBO formulation for the Sharpe Ratio maximization as a building block for the complete optimization model.

Therefore, we do not emphasize the computational time required to obtain the solutions as it is not the primary focus of our study. Instead, we draw attention to the quality of the results in terms of objective function value, which constitutes a larger interest from a business perspective with respect to an investigation of the time-wise scaling properties of the approach.

\subsection{$\lambda$ Hyperparameter Optimization}

To ensure the generation of high-quality solutions, we conduct a calibration procedure to determine the optimal values for $\lambda_0$, $\lambda_1$ and $\lambda_2$ in our formulations, separately for each of the goals of our investigation. The objective of this calibration process is to study how the Sharpe Ratio and Diversification measures differ for varying values of the hyperparameters, as well as to create a significant energy gap between feasible and infeasible solutions within the QUBO matrix. This gap facilitates the identification of correct and incorrect results in terms of constraint satisfaction. Additionally, the calibration aims to establish a correlation where higher-quality portfolios are associated with lower energy values, enabling the optimization process to produce improved solutions.

Regarding the analysis on the sole Sharpe Ratio maximization, we conducted multiple QUBO instances with varying parameters and repeated the procedure for each combination of QUBO formulation and solver. Through this process, we obtained the following optimal values. A detailed explanation of our analysis can be found in the Appendix. The optimal values we found are as follows:
\begin{itemize}
    \item For the Sharpe Ratio Proxy formulation, solving with qbsolv: $\lambda_0 = 1.2631$, $\lambda_1 = 300$.
    \item For the Sharpe Ratio Proxy formulation, solving with D-Wave Leap: $\lambda_0 = 1.2631$, $\lambda_1 = 300$.
    \item For the Proposed Sharpe Ratio Formulation, solving with qbsolv: $\lambda_0 = 0.7$, $\lambda_1 = 300$.
    \item For the Proposed Sharpe Ratio Formulation, solving with D-Wave Leap: $\lambda_0 = 0.7$, $\lambda_1 = 300$.
\end{itemize}

\subsection{Results}\label{results}
\begin{figure}[t!]\label{errorplot_fig}
    \centering
    \resizebox{\textwidth}{!}{\includegraphics{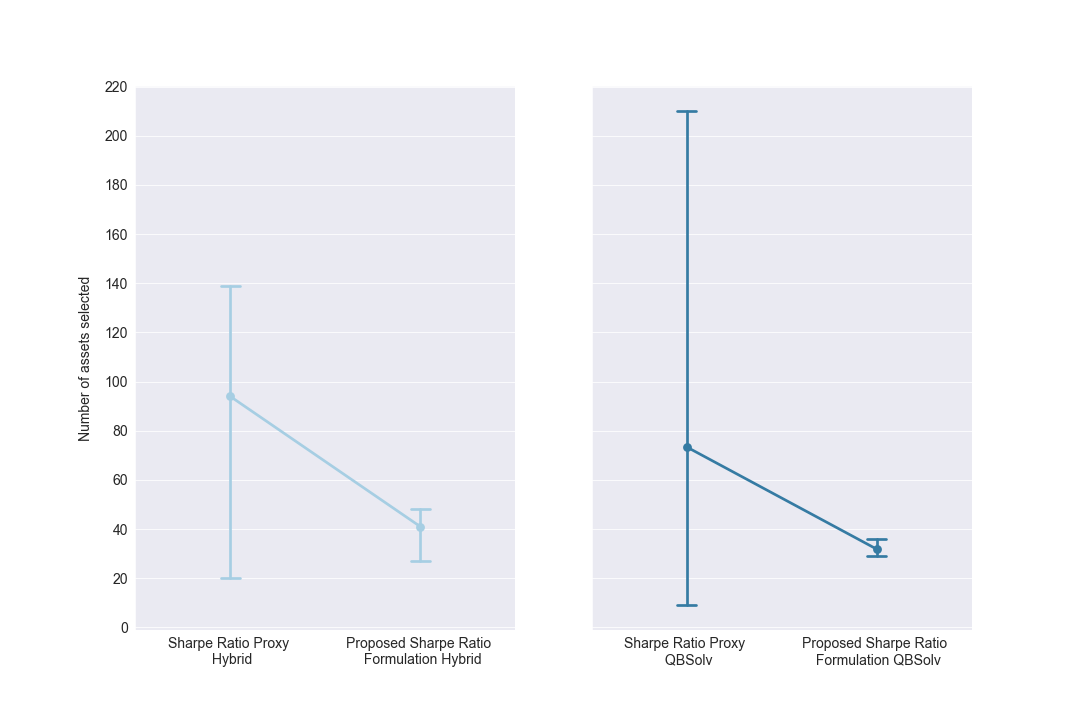}}
    \caption{Minimum, maximum and mean (identified by the dot) in the number of assets selected over the 10 feasible solutions found. Left plot reports the statistics solving the two QUBO formulations using D-Wave Hybrid. Right plot, likewise, shows these results derived from the QBSolv solver.} 
    \label{output}
\end{figure}

Figure \ref{output} presents statistics regarding the number of assets selected using the two formulations and different solvers. With the optimization strategy fixed, the solutions exhibit a consistent pattern: the Proposed Sharpe Ratio Formulation results in a decrease in both the mean number of assets selected and the variability. This behavior is observed across all solvers and can be attributed to the differences in the number of variables and block sizes between the two QUBO formulations, as discussed previously.
\begin{figure*}[t!]\label{sr_diversif}
    \centering
    {
    \includegraphics[width=1\textwidth]{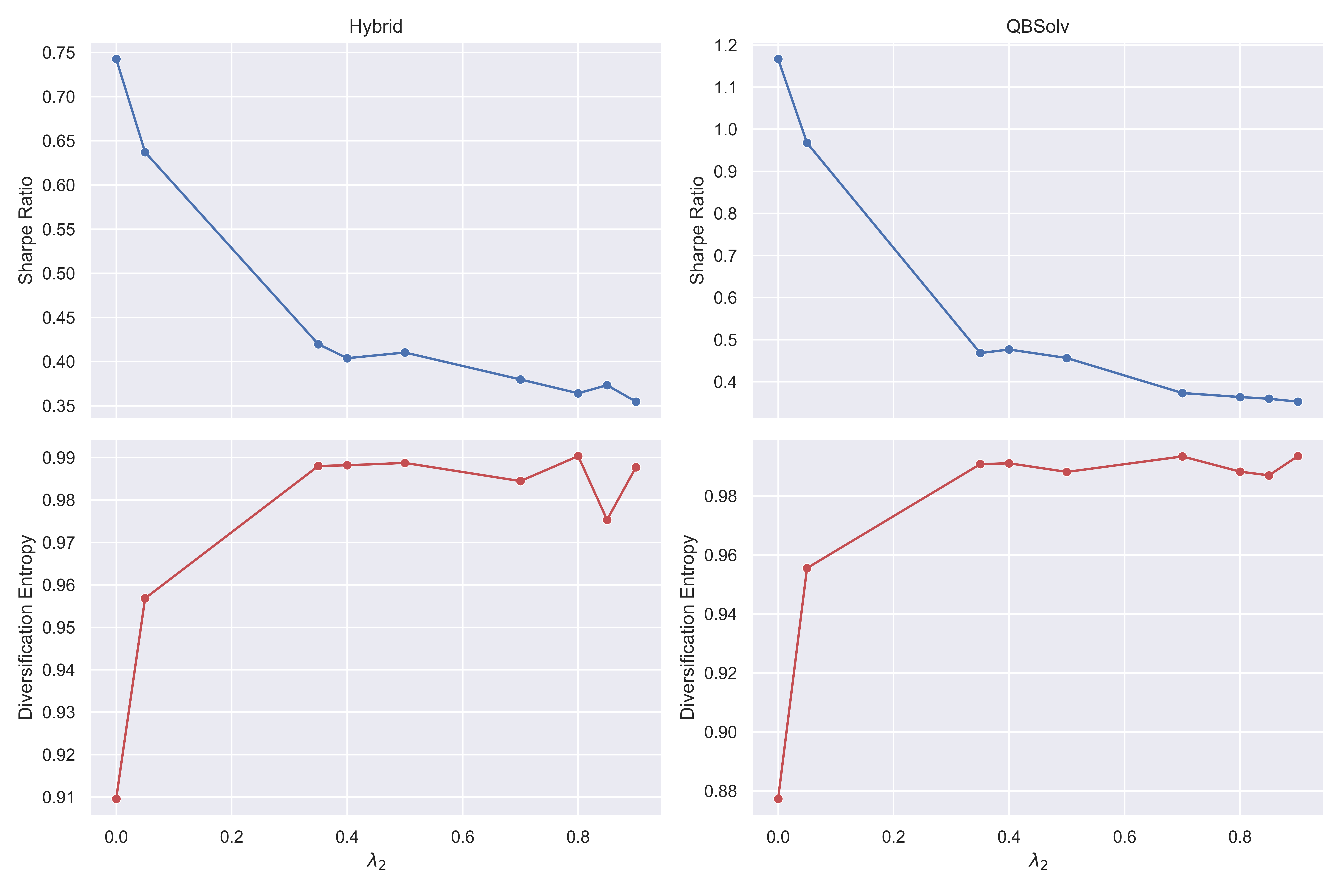}
    \caption{Optimization runs on D-Wave's QBSolv (CPU) and D-Wave's Hybrid (QPU) of multiple QUBO instances as $\lambda_2$ varies. The plots reports the behavior of Sharpe Ratio and Diversification Entropy, respectively, as $\lambda_2$ increases. All runs are feasible, considering the constraint $\mu^Ty = 1$ satisfied if $\mu^Ty$ is in a neighbourhood of 1 (ref. to Section \ref{proposed_form}), up to a factor equal to $2.5*10^-4$, which is given by multiplying the minimum discretization coefficient by the minimum expected return.} 
    \label{diversif_results}}
\end{figure*}

\begin{figure*}[t!]\label{max_sharpe_qubo_comparison}
    \centering
    \resizebox{\textwidth}{!}
    {\includegraphics{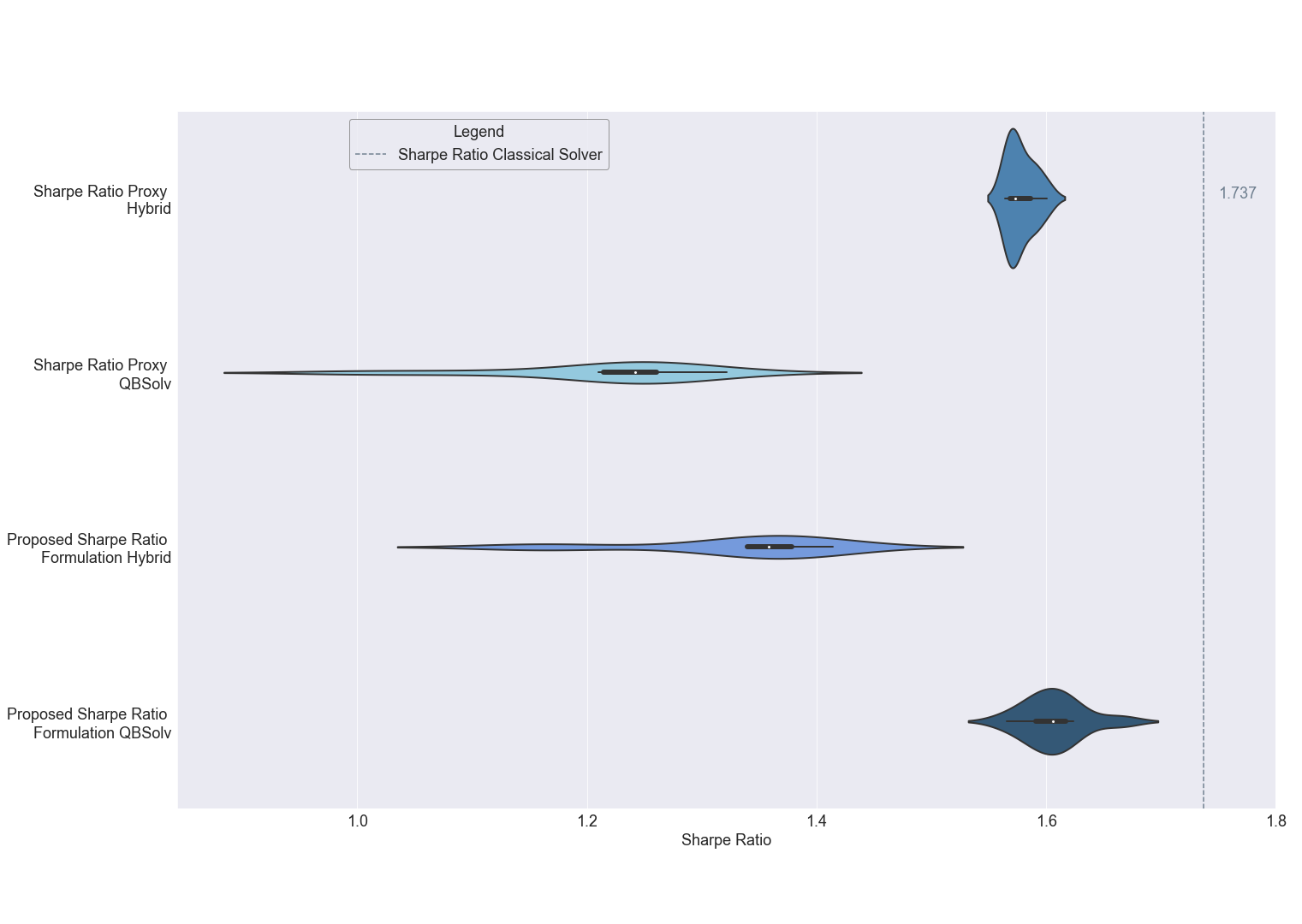}}
    \caption{Results provided by each combination of QUBO formulation and solver. The statistics are drawn from $10$ feasible solutions with fixed values for the $\lambda$ coefficients. All solutions are feasible: for the Sharpe Ratio Proxy formulation, feasibility is given by the sum of asset weights equal to $1$, while for the Proposed Sharpe Ratio Formulation we consider the constraint satisfied if $\bm{\mu}^T\bm{y}$ is in a neighbourhood of $1$, up to a factor equal to $2.5*10^{-4}$.} 
    \label{fig_results}
\end{figure*}

Figure \ref{diversif_results} shows the Sharpe Ratio and Diversification measures of the optimized portfolio as the hyperparameters vary. For higher values of $\lambda_2$, whose scope is to induce solutions with a higher degree of diversification, we report results in accordance to the expectations. As the diversification term impacts more and more significantly the optimization, the Sharpe Ratio value tends to decrease: this is due to the optimization being decreasingly incentivized to maximize the Sharpe Ratio, favoring solutions that allocate investments over different sectors but having less impact on the expected return or covariance of the assets. The runs have been performed having fixed $\lambda_0 = 0.44$ and $\lambda_1 = 10000$. A value high enough for $\lambda_1$ ensures that the constraint expressed in Equation \ref{proposed_constr} is met throughout all the runs. Then, for our scope, the exact value of $\lambda_0$ is not as relevant as the actual ratio between $\lambda_0$ and $\lambda_2$, which determine the magnitude of impact of one measure over the other (respectively, Sharpe Ratio and Diversification).

In our investigation on the appropriateness of our Sharpe Ratio Maximization formulation, where hence the diversification term is discarded, once the optimal values for $\lambda_0$ and $\lambda_1$ are fixed, we retrieve 10 additional feasible solutions for each combination of QUBO formulation and solver. We gather statistics on the results in terms of Sharpe Ratio values. In Figure \ref{fig_results}, we compare these results with the solution obtained by solving the Max-Sharpe problem implemented in the PyPortfolioOpt library \cite{pypfopt}.

The Sharpe Ratio Proxy and the Proposed Sharpe Ratio Formulation result in 3888 and 5184 binary variables, respectively. Among the QUBO solutions, the best results are achieved by solving the Proposed Sharpe Ratio Formulation with qbsolv. It is worth noting that, for both formulations, the best performances are obtained by different solvers: D-Wave Hybrid in one case and qbsolv in the other. This discrepancy can be attributed not only to the different number of variables but also to the specific patterns of the QUBO matrices. With 432 assets, the block size is 9 variables for the Sharpe Ratio Proxy formulation and 12 variables for the Proposed Sharpe Ratio Formulation.

This difference in block size may influence the behavior of the solvers and lead to varying results. The gap between the classical and the best QUBO solutions can be attributed to the capabilities of the solvers. A finer discretization would allow for a closer representation of continuous values but may result in an increase in the number of variables, potentially affecting the solver's performance and leading to suboptimal results.

\section{Conclusions}\label{sec:conclusions}

The Portfolio Optimization problem is a well-known task in Financial Services and has recently drawn attention within the quantum computing literature thanks to the applicability of quantum annealers to solve the problem.

In this work we tackle a specific strategy to find the optimal allocation of investments over a set of assets, namely the Sharpe Ratio maximization, while optimizing a diversification measure that allows to spread investments across multiple sectors, leading to a non-trivial optimization task. When modeling the Sharpe Ratio, the first building block of our work, we extend a QUBO formulation proposed by \cite{venturelli}, highlighting its benefits and potential drawbacks and propose a novel QUBO formulation to address the drawbacks. Then, we formulate the complete QUBO by taking into account both measures of portfolio quality. We run our experiments on classical and quantum computing hardware and elaborate on the results both in terms of the QUBO formulation and in terms of the optimization solver. Future works might include a deeper investigation of different solvers, the corresponding capabilities of the Proposed Sharpe Ratio Formulation and the formulation of additional common needs from the Financial Services industry.

\vspace{6pt}

\bmhead{Acknowledgments}
{The authors wish to thank Marco Magagnini, Data Reply S.r.l., for fruitful discussions and Massimiliano Incudini, Università degli Studi di Verona, for useful suggestions.}

\newpage





\appendix

\section*{\appendixname\,}

\subsection*{Calibration Analysis}

We analyse in further details the calibration procedure that led to the choice of the optimal values for the coefficients $\lambda_0$ and $\lambda_1$ described in Section \ref{results}.

\begin{figure}[h!]
\captionsetup[subfigure]{justification=centering}
    \centering
    \begin{subfigure}{0.48\textwidth}
    \resizebox{\textwidth}{!}
        {\includegraphics{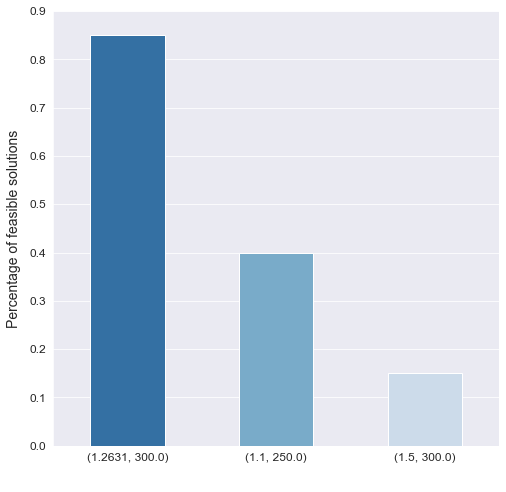}}
        \caption{Sharpe Ratio Proxy Hybrid}
    \end{subfigure}
    \begin{subfigure}{0.48\textwidth}
        \resizebox{\textwidth}{!}
        {\includegraphics{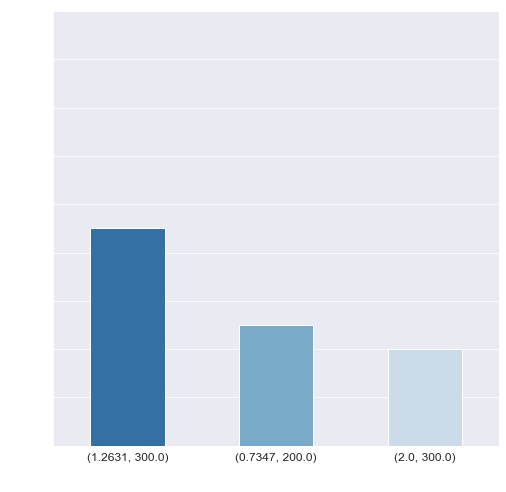}}
        \caption{Sharpe Ratio Proxy QBSolv}
    \end{subfigure} \\
    \bigskip
    \begin{subfigure}{0.48\textwidth}
        \resizebox{\textwidth}{!}
        {\includegraphics{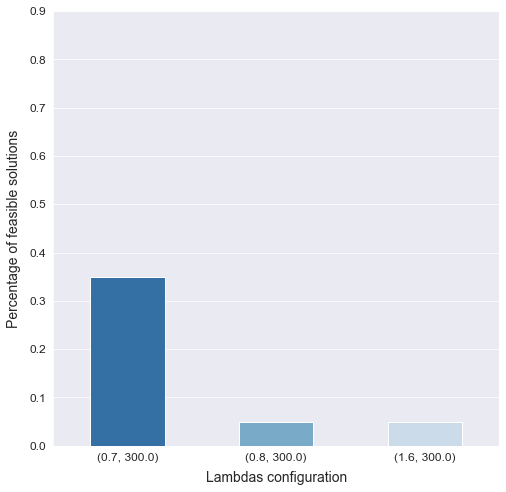}}
        \caption{Proposed Sharpe Ratio \\ Formulation Hybrid}
    \end{subfigure}
    \begin{subfigure}{0.48\textwidth}
        \resizebox{\textwidth}{!}
        {\includegraphics{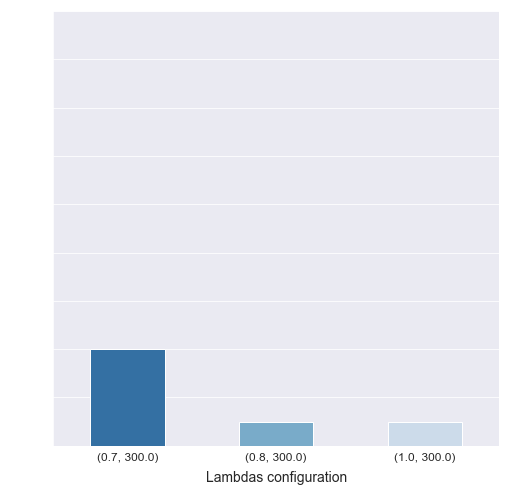}}
        \caption{Proposed Sharpe Ratio \\Formulation QBSolv}
    \end{subfigure}
    \bigskip
    \caption{Percentage of feasible solutions obtained from 20 runs for each lambda configuration, for each combination of formulation and solver.}
    \label{appendix_fig}
\end{figure}

In Figure \ref{appendix_fig} we show the results of our approach: for each combination of formulation and solver, we initially tested a grid of values for both coefficients; once the best lambda pairs were identified in terms of Sharpe Ratio value and corresponding constraint value, we performed 20 runs on each of these pairs and finally selected as optimal the one with the highest percentage of feasible solutions. Figure \ref{appendix_fig} reports multiple configurations of lambdas and the corresponding percentage number of feasible solutions. Configurations with lower $\lambda_0$ and higher $\lambda_1$ with respect to those represented here would, on one hand, lead to a higher percentage, while on the other decrease the value of Sharpe Ratio of the corresponding solutions. Therefore from Figure \ref{appendix_fig} stem the configurations that aim at maximizing the Sharpe Ratio, but also provide a number of feasible solutions as high as possible.

Multiple additional configurations have been tested, which however only led to infeasible solutions. Namely:
\begin{itemize}
    \item Sharpe Ratio Proxy Hybrid:
    \begin{itemize}
        \item $\lambda_0 = 2.4 $ , $\lambda_1 = 200 $
        \item $\lambda_0 = 2 $ , $\lambda_1 = 200 $
        \item $\lambda_0 = 2.4 $ , $\lambda_1 = 300 $
        \item $\lambda_0 = 2.6 $ , $\lambda_1 = 300 $
    \end{itemize}
    \item Sharpe Ratio Proxy QBSolv:
    \begin{itemize}
        \item $\lambda_0 = 3.5 $ , $\lambda_1 = 300 $
        \item $\lambda_0 = 2.6 $ , $\lambda_1 = 300 $
    \end{itemize}
    \item Proposed Sharpe Ratio Formulation Hybrid:
    \begin{itemize}
        \item $\lambda_0 = 2.6 $ , $\lambda_1 = 300 $
    \end{itemize}
    \item Proposed Sharpe Ratio Formulation QBSolv:
    \begin{itemize}
        \item $\lambda_0 = 50 $ , $\lambda_1 = 500 $
        \item $\lambda_0 = 20 $ , $\lambda_1 = 500 $
        \item $\lambda_0 = 20 $ , $\lambda_1 = 700 $
        \item $\lambda_0 = 20 $ , $\lambda_1 = 1000 $
        \item $\lambda_0 = 0.9474 $ , $\lambda_1 = 300 $

    \end{itemize}
\end{itemize}

\newpage

\end{document}